\documentclass[sigconf]{acmart}
\usepackage{multirow}
\usepackage{balance}

\AtBeginDocument{%
  \providecommand\BibTeX{{%
    \normalfont B\kern-0.5em{\scshape i\kern-0.25em b}\kern-0.8em\TeX}}}

\copyrightyear{2024}
\acmYear{2024}
\setcopyright{acmlicensed}
\acmConference[MRAC '24] {Proceedings of the 2nd International Workshop on Multimodal and Responsible Affective Computing}{November 1, 2024}{Melbourne, VIC, Australia.}
\acmBooktitle{Proceedings of the 2nd International Workshop on Multimodal and Responsible Affective Computing (MRAC '24), November 1, 2024, Melbourne, VIC, Australia}
\acmISBN{979-8-4007-1203-6/24/11}
\acmDOI{10.1145/3689092.3689407}

\begin{document}

\title{Improving Multimodal Emotion Recognition by Leveraging Acoustic Adaptation and Visual Alignment}



\author{Zhixian Zhao}
\affiliation{%
  \institution{\mbox{Northwestern Polytechnical University}}
  \city{Xi’an}
  \country{China}}
\email{zxzhao@mail.nwpu.edu.cn}

\author{Haifeng Chen}
\affiliation{%
  \institution{Shaanxi University of Science and Technology}
  \city{Xi’an}
  \country{China}}
\email{chenhaifeng@sust.edu.cn}

\author{Xi Li}
\affiliation{%
  \institution{Shaanxi University of Science and Technology}
  \city{Xi’an}
  \country{China}}
\email{lixii@sust.edu.cn}

\author{Dongmei Jiang}
\affiliation{%
  \institution{Peng Cheng Laboratory}
  \city{Shenzhen}
  \country{China}}
\affiliation{%
  \institution{\mbox{Northwestern Polytechnical University}}
  \city{Xi’an}
  \country{China}} 
\email{jiangdm@nwpu.edu.cn}

\author{Lei Xie}
\authornote{Corresponding author.}
\affiliation{%
  \institution{\mbox{Northwestern Polytechnical University}}
  \city{Xi’an}
  \country{China}
}
\email{lxie@nwpu.edu.cn}
\renewcommand{\shortauthors}{Zhixian Zhao, Haifeng Chen, Xi Li, Dongmei Jiang, \& Lei Xie}
\begin{abstract}
Multimodal Emotion Recognition (MER) aims to automatically identify and understand human emotional states by integrating information from various modalities. However, the scarcity of annotated multimodal data significantly hinders the advancement of this research field. This paper presents our solution for the MER-SEMI sub-challenge of MER 2024. First, to better adapt acoustic modality features for the MER task, we experimentally evaluate the contributions of different layers of the pre-trained speech model HuBERT in emotion recognition. Based on these observations, we perform Parameter-Efficient Fine-Tuning (PEFT) on the layers identified as most effective for emotion recognition tasks, thereby achieving optimal adaptation for emotion recognition with a minimal number of learnable parameters. Second, leveraging the strengths of the acoustic modality, we propose a feature alignment pre-training method. This approach uses large-scale unlabeled data to train a visual encoder, thereby promoting the semantic alignment of visual features within the acoustic feature space. Finally, using the adapted acoustic features, aligned visual features, and lexical features, we employ an attention mechanism for feature fusion. On the MER2024-SEMI test set, the proposed method achieves a weighted F1 score of 88.90\%, ranking fourth among all participating teams, validating the effectiveness of our approach.
\end{abstract}





\maketitle
\section{Introduction}

Automatic emotion recognition is crucial for human-computer interaction (HCI), allowing computers to detect and respond to users' emotional states~\cite{Intro, Intro_ER_ACM}. Multimodal emotion recognition (MER), based on supervised learning, has shown promising results~\cite{MER1, MER2, MER3}. However, the annotation process is costly and time-consuming, limiting the scale of labeled MER datasets and hindering performance. The MER-SEMI challenge, a sub-challenge of the Multimodal Emotion Recognition Challenge \cite{mer2023, mer2024}, addresses this issue by providing a labeled emotional dataset alongside substantial unlabeled data, enabling participants to explore effective unsupervised or semi-supervised learning strategies.

Pre-trained transformer models \cite{hubert, wav2vec2.0, whisper, data2vec2} have achieved notable success across various speech tasks, excelling in capturing phonetic structures, temporal dependencies, and acoustic features. Studies \cite{different_layer1, different_layer2} suggest that essential task-specific information may reside in the hidden representations of different transformer layers, yet there is limited exploration of their contributions to speech emotion recognition. Typically, current parameter fine-tuning methods \cite{SER_fintune_whisper, wang2022finetunedwav2vec20hubertbenchmark, SER_adapterfinetunewith_icassp24} involve modifying the entire model, which overlooks the varying significance of features across layers. To improve the performance of pre-trained speech models in emotion recognition, it is beneficial to incorporate adapters when fine-tuning certain intermediate layers. This approach aligns features with emotion recognition requirements, reduces training parameters, and preserves the model’s generalization capability.

The visual modality provides rich non-verbal information, such as facial expressions, body language, and gestures, essential for computer vision and natural language processing tasks \cite{vision1, vision2}. To fully leverage multimodal information, comparative cross-modal pretraining methods \cite{Contra_Learning1, Contra_Learning2, Contra_Learning_gemo-clip} have been developed. For instance, CLIP \cite{clip} achieves a shared semantic space for visual and textual understanding by jointly training image and text encoders. Baseline results \cite{mer2024} indicate that the visual modality's performance in emotion recognition is relatively weaker compared to the acoustic modality. Therefore, after obtaining optimal acoustic features, the visual modality can be aligned with the acoustic feature space. Contrastive learning methods can establish relationships between acoustic and visual modalities, enhancing the visual modality's ability to capture emotional information more accurately.

\begin{figure*}[t]
\centering
\includegraphics[width=0.95\textwidth]{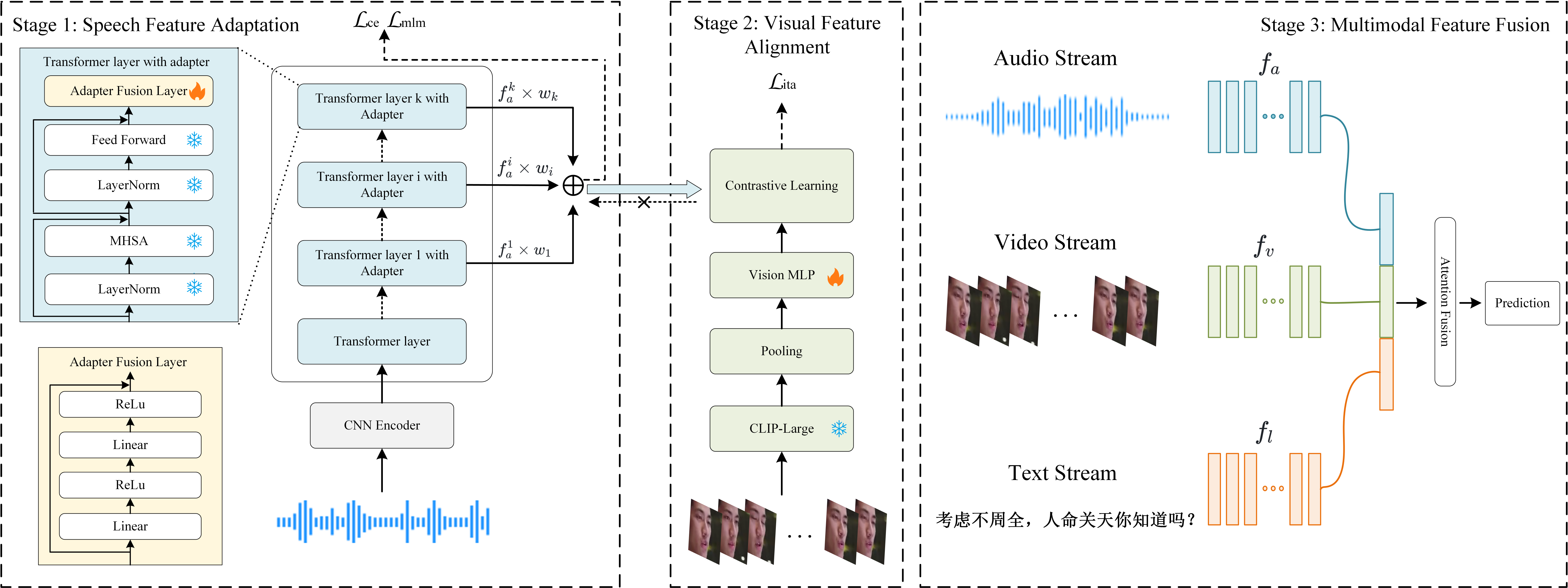}
\caption{The proposed multimodal emotion recognition model framework}
\label{fig:1}
\vspace{-8pt}
\end{figure*}

Based on the above discussions, we propose a semi-supervised multimodal emotion recognition method comprising three stages: acoustic feature adaptation, visual feature alignment, and multimodal feature fusion. The main flow is shown in Figure \ref{fig:1}. First, to optimize acoustic modality features for emotion recognition, we conducted an empirical study exploring the performance of different layers of the HuBERT-large model \cite{hubert} and the effectiveness of multi-layer feature fusion. Guided by these empirical findings, we propose a simple and effective parameter-efficient fine-tuning method. This method enhances recognition performance by incorporating adapters into well-performing intermediate transformer layers and dynamically fusing hidden representations across these layers using learnable weights. Second, to enhance the emotional representation capability of visual modality features, we perform contrastive learning between the fine-tuned acoustic features and visual features processed by a multilayer perceptron (MLP). We leverage a substantial amount of unlabeled visual and audio data to pretrain the vision MLP in an unsupervised manner, ensuring the visual features adapt to the acoustic feature space. Finally, we employ an attention-based feature fusion module to integrate the acoustic, visual, and textual features, achieving a weighted F1 score of 88.90\% on the test set.

\section{Method}

\subsection{Acoustic Feature Adaptation}
\label{SFA}

Based on the findings of the baseline study \cite{mer2024}, we select the HuBERT-large model \cite{hubert}, which has demonstrated superior performance in emotion recognition tasks, as the feature extractor. Since pre-trained transformer models (e.g., HuBERT and wav2vec2.0 \cite{wav2vec2.0}) capture unique hidden representations across different layers during audio processing and that these layers may contain complementary information, we utilize the temporal pooling features ($f^{i}_{a} \in \mathbb{R}^{\mathrm{d}_{\mathrm{a}}}, i \in {1,2,...,k}$) from $k$ consecutive middle layers of the HuBERT-large model for the emotion recognition task.

To adapt these features more effectively to the emotion recognition task while maintaining the generalization capability of the pre-trained model, we introduce adapters in these $k$ transformer layers for efficient parameter fine-tuning. The implementation details of the adapters are depicted in Figure \ref{fig:1}(Stage 1). Each adapter consists of a bottleneck structure, including a dimension reduction projection layer that reduces the hidden dimension from $\mathrm{d}_{\mathrm{a}}$ to $\mathrm{\hat{d}}$, followed by a ReLU non-linear activation function, and then an up-projection layer that restores the dimension back to $\mathrm{d}_{\mathrm{a}}$. Given an input feature $x$, the output $y$ of the adapter can be represented as:
\begin{equation}
y = x + \mathrm{ReLU}(W_{up} \mathrm{ReLU}(W_{down} x + b_{down}) + b_{up}),
\end{equation}
where $W_{down} \in \mathbb{R}^{\mathrm{d}_{\mathrm{a}} \times \mathrm{\hat{d}}}$, $W_{up} \in \mathbb{R}^{\mathrm{\hat{d}} \times \mathrm{d}_{\mathrm{a}}}$, $b_{down} \in \mathbb{R}^{\mathrm{\hat{d}}}$, $b_{up} \in \mathbb{R}^{\mathrm{d}_{\mathrm{a}}}$ are trainable parameters.

To account for the varying contributions of different layers to the emotion recognition task, we introduce learnable weights ($w_i, i \in \{1,2,...,k\}$) to fuse the output features of these $k$ transformer layers, resulting in the fused feature ${f}^{fusion}_{a}= {\textstyle \sum_{i=1}^{k}} w_i \cdot {f}^{i}_{a}$, where ${f}^{fusion}_{a}\in \mathbb{R}^{\mathrm{d}_{\mathrm{a}}}$. We use two types of loss functions to optimize the model. The first is an unsupervised masked reconstruction loss, inspired by \cite{hubert}, this approach predicts the masked portions of the acoustic features using contextual information to learn more robust acoustic representations. We generate masked features ${f}^{masked}_{a} \in \mathbb{R}^{ \mathrm{d}_{\mathrm{a}}}$ through a multi-layer perceptron (MLP) consisting of two fully connected layers and a ReLU layer, and measure the difference between the original and masked features using the mean squared error (MSE) loss function. The second loss function is a supervised cross-entropy (CE) loss. After passing the masked features through a fully connected layer to obtain the predicted results $x^{pred}$, we use the CE loss function to calculate the loss between the predicted results and the labels $y^{label}$. These two loss functions can be expressed as:
\begin{equation}
\mathcal{L}_{ce}=CE(x^{pred}, y^{label}),
\end{equation}
\begin{equation}
\mathcal{L}_{mlm}=MSE({f}^{masked}_{a}, {f}^{fusion}_{a}),
\end{equation}
Thus, the overall objective function for fine-tuning the model is defined as:
\begin{equation}
\mathcal{L}=\mathcal{L}_{ce} + \mathcal{L}_{mlm}.
\end{equation}

\vspace{-8pt}
\subsection{Visual Feature Alignment}
\label{VFA}

Due to the semantic disparity between the visual features extracted by CLIP-large \cite{clip} and the audio features, we perform pre-training feature alignment for the visual modality prior to multimodal feature fusion, as illustrated in Figure \ref{fig:1}(Stage 2). Inspired by \cite{LLaVA}, we first pre-train a vision MLP to align visual representations with the semantic space of pre-trained speech emotion representations. During this process, the weights of the CLIP-large model and the fine-tuned HuBERT-large model are kept frozen, and only the weights of the vision MLP are trained.

We introduce an video-audio contrastive learning method. An input video $I$ is encoded by the CLIP-large model and then average-pooled to obtain $f_{\hat{v}} \in \mathbb{R}^{\mathrm{d}_{\mathrm{v}}}$. Then, an MLP maps the visual embedding to the same dimensionality as the audio embedding, resulting in mapped visual embedding $f{v} \in \mathbb{R}^{\mathrm{d}_{\mathrm{a}}}$. This transformation can be represented as:
\begin{equation}
f_{v} = \mathrm{ReLU}(W_{1} \mathrm{ReLU}(W_{2}f_{\hat{v}} + b_{1}) + b_{2}),
\end{equation}
where $W_1 \in \mathbb{R}^{d_v \times \mathrm{d}_{\mathrm{a}}}$ and $W_2 \in \mathbb{R}^{\mathrm{d}_{\mathrm{a}} \times \mathrm{d}_{\mathrm{a}}}$, $b_1 \in \mathbb{R}^{\mathrm{d}_{\mathrm{a}}}$ and $b_2 \in \mathbb{R}^{\mathrm{d}_{\mathrm{a}}}$ are trainable parameters of the vision MLP. The goal of the MLP is to align the visual and audio features so they can operate within a unified feature space.

To compute the similarity between the visual and audio embeddings, we first use the fine-tuned HuBERT-large model to convert the input audio $A$ into an embedding sequence $f_a \in \mathbb{R}^{\mathrm{d}_{\mathrm{a}}}$. The video-audio similarity can be defined as: $s(I, A) = f_{v}^\top f_{a} / \left \| f_{v} \right \| \cdot \left \| f_{a} \right \|$ and the audio-video similarity as: $s(A,I) = f_{a}^\top f_{v} / \left \| f_{a} \right \| \cdot \left \| f_{v} \right \|$. For each video and audio pair, we calculate the normalized softmax similarity scores to measure the similarities from video to audio and from audio to video:
\begin{equation}
p_{j}^{\mathrm{i2a}}(I) = \frac{\exp(s(I_j, A)/\tau)}{\sum_{j=1}^J \exp(s(I_j, A)/\tau)}, \quad p_{j}^{\mathrm{a2i}}(A) = \frac{\exp(s(A_j,I)/\tau)}{\sum_{j=1}^J \exp(s(A_j,I)/\tau)},
\end{equation}
where $\tau$ is a learnable temperature parameter. Let $y^{\mathrm{i}2\mathrm{a}}$ and $y^{\mathrm{a}2\mathrm{i}}$ represent the ground truth one-hot encoded similarities, with a probability of 0 for mismatched video-audio pairs and 1 for matched pairs. The video-audio contrastive loss $\mathcal{L}_{\mathrm{ita}}$ is defined as the cross-entropy $\mathrm{H}$ between $p$ and $y$:
\begin{equation}
\mathcal{L}_{\mathrm{ita}} = \frac{1}{2} \mathbb{E}_{(I,A)\thicksim D} \left[ \mathrm{H}(y^{\mathrm{i}2\mathrm{a}}(I), p^{\mathrm{i}2\mathrm{a}}(I)) + \mathrm{H}(y^{\mathrm{a}2\mathrm{i}}(A), p^{\mathrm{a}2\mathrm{i}}(A)) \right].
\end{equation}

\vspace{-8pt}
\subsection{Multimodal Feature Fusion}

Before conducting feature fusion, we outline the extraction processes for the three modalities. Acoustic Features: we utilize the fine-tuned Chinese-HuBERT-Large model \cite{hubert} to extract the speech representation $f_a$ for each audio sample. Visual Features: we input the facial images extracted using the OpenFace toolkit \cite{openface} into the CLIP-large model \cite{clip} to extract the visual features. Subsequently, we use the vision MLP introduced in Section \ref{VFA} as the feature extractor to obtain the feature representation $f_v$. Lexical Features: we use the Baichuan2 \cite{Baichuan2} model to extract the feature representation $f_l$ from the checked transcripts of the  Train\&Val set, as well as the subtitle files of the unlabeled data.


After obtaining the feature vectors for the three modalities $f_m \in \mathbb{R}^{\mathrm{d}_{\mathrm{m}}}$, $m \in (a,v,l)$,  we use an MLP composed of several fully connected layers and ReLU activation functions to map each modality's features to the same dimension. Considering the varying importance of each modality, we stack the three modality features together and calculate the attention score $\alpha$ for each modality:
\begin{equation}
h = \mathrm{Concat}(h_{a}, h_{l}, h_{v}),
\end{equation}
\begin{equation}
\alpha = \mathrm{softmax}(h^\top W_{\alpha} + b_{\alpha}),
\end{equation}
where $W_{\alpha}$ and $b_{\alpha}$ are trainable parameters. The final fused feature is given by $z = h \alpha$.

\vspace{-3pt}
\section{Experiments and Results}
\subsection{Dataset and Implementation Details}

\begin{table}[]
  \caption{Statistics of the MER 2024 dataset}
  \label{tab:mer24}
\resizebox{0.9\linewidth}{!}{
\begin{tabular}{lccc}
\hline
Dataset & Labeled & Unlabeled   & Duration (hr:min:sec)  \\ \hline
Train\&Val   & 5030    & 0           & 05:56:39  \\ \hline
MER-SEMI     & 0       & 1169/115595 & 100:38:49 \\
MER-NOISE    & 0       & 1170/115595 & 100:38:49 \\ \hline
\end{tabular}}
\vspace{-10pt}
\end{table}

\textbf{Dataset:} We conduct experiments using the MER-SEMI dataset, as detailed in Table \ref{tab:mer24}. The Train\&Val set consists of 5,030 video clips with both discrete and dimensional emotion labels. Given the absence of a predefined training/validation split, we utilize five-fold cross-validation on the Train\&Val set \cite{mer2024}. To evaluate model generalization, the MER-SEMI track includes 1,169 unlabeled video clips in a test set, drawn from a total of 115,595 unlabeled data. Participants must predict discrete emotion labels across all unlabeled data, not just the test set. The set of discrete emotion labels includes six categories: \textit{neutral}, \textit{anger}, \textit{happiness}, \textit{sadness}, \textit{worry}, and \textit{surprise}. We use a weighted average F1 score as the evaluation metric, aligning with the official baseline.

\textbf{Implementation Details:} For acoustic features adaptation, we select the 16th to 21st layers (total \( k=6 \) layers) of the HuBERT-large model and incorporate adapters into these layers for fine-tuning (refer to Section \ref{empirical_study}). The feature dimension $\mathrm{\hat{d}}$ within the Adapter is set to 128. During fused feature computation, we assign an initial weight of 1.0 to the 18th layer for its optimal performance, while the other layers are initially weighted at 0.0. Pre-training is conducted on the Train\&Val set with a batch size of 16 and a learning rate of 1e-4. We use the Adam optimizer with a weight decay of 0.02. For visual feature alignment, the vision MLP is trained on unlabeled data with a batch size of 1024 and a learning rate of 1e-4. For the feature fusion module, features from the three modalities are mapped to 256 dimensions through an MLP for fusion, with a learning rate set to 1e-4. The dimensions of the visual, audio, and lexical features are \( \mathrm{d}_{\mathrm{v}} = 768 \), \( \mathrm{d}_{\mathrm{a}} = 1024 \), and \( \mathrm{d}_{\mathrm{l}} = 5120 \), respectively.

\vspace{-5pt}
\subsection{Result and Discussion}
\subsubsection{Empirical study}
\label{empirical_study}

\begin{figure}[t]
\centering
\includegraphics[width=0.9\linewidth]{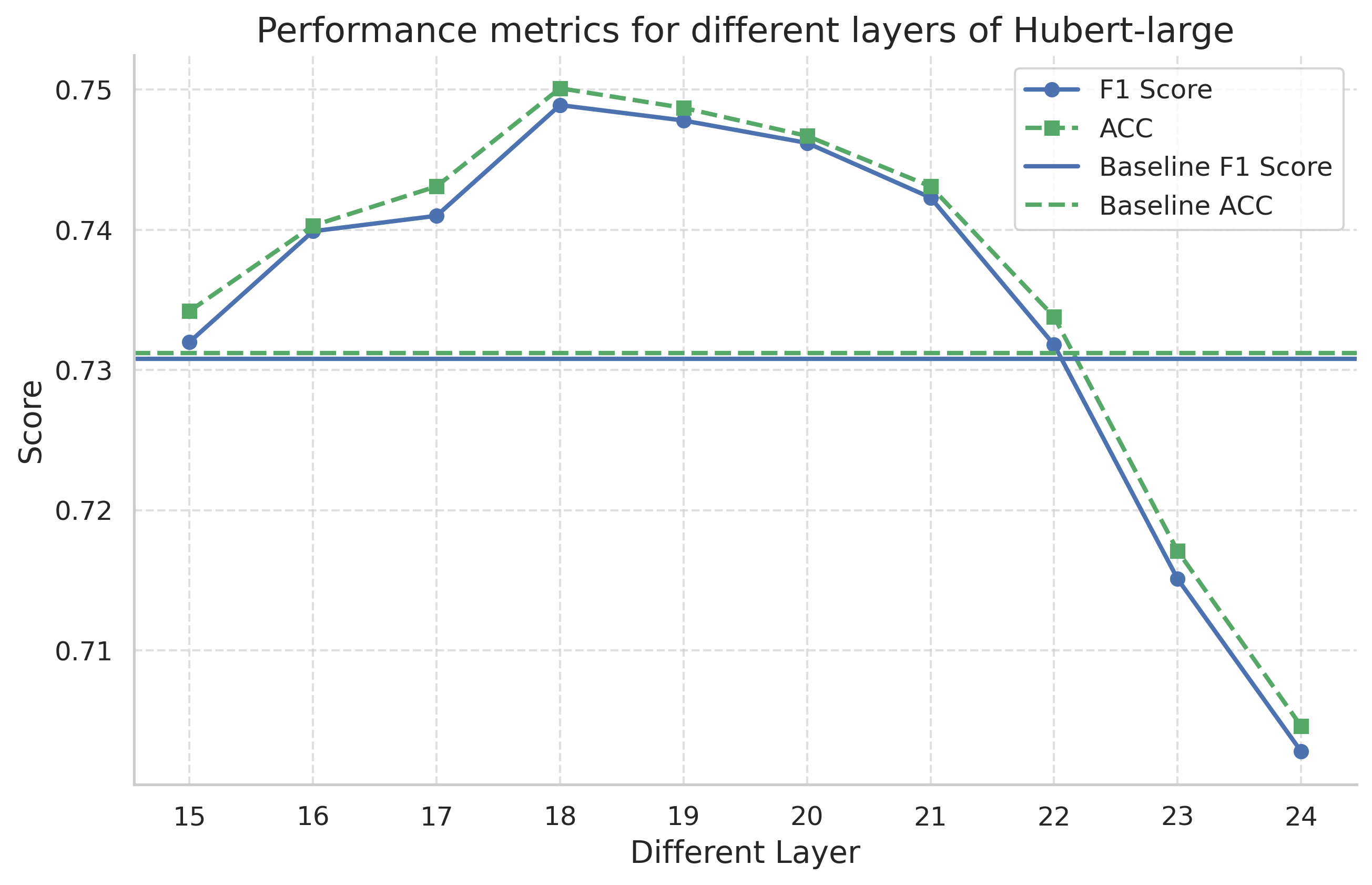}
\caption{Comparison of the performance of features from different layers of HuBERT-large.}
\vspace{-8pt} 
\label{fig:2}
\end{figure}

To validate the performance of features from different transformer layers in emotion recognition tasks, we extract the features from the last 10 layers of the HuBERT-large model and evaluate their performance on the Train\&Val set, as shown in Figure \ref{fig:2}. The rationale for focusing on the last 10 layers is that shallow layers typically encode low-level features, such as pitch and short-term energy fluctuations, while deeper layers are more adept at capturing high-level features and global semantic information, which are crucial for emotion recognition. As shown in the figure, features from the 18th layer demonstrate the best performance, significantly surpassing the baseline score, indicating that this layer captures more representative emotional features. Furthermore, features from the 16th through 21st layers consistently outperform the baseline, indicating that these intermediate layers are more suitable for emotion recognition tasks. These layers balance capturing both low-level and high-level information, providing rich audio content while minimizing noise interference. This finding is of significant importance for subsequent model fine-tuning efforts.

\begin{table}[]
\caption{Evaluation of Unimodal and Multimodal Feature Performance.}
\label{table:2}
\resizebox{0.95\linewidth}{!}{
\begin{tabular}{lcccc}
\hline
\multicolumn{3}{c|}{Features}                                                        & \multirow{2}{*}{\begin{tabular}[c]{@{}c@{}}Train\&Val\\ F1-Score(↑)\end{tabular}} & \multirow{2}{*}{\begin{tabular}[c]{@{}c@{}}MER-SEMI\\ F1-Score(↑)\end{tabular}} \\ 
\multicolumn{1}{c}{A} & V                           & \multicolumn{1}{c|}{T}         &                                                                                   &                                                                                 \\ \hline
\multicolumn{5}{c}{Unimodal Results}                                                                                                                                                                                                                       \\ \hline
HL                    & -                           & \multicolumn{1}{c|}{-}         & 72.82±0.30                                                                        & 83.49                                                                           \\
HL(18)                & -                           & \multicolumn{1}{c|}{-}         & 74.40±0.46                                                                        & 84.78                                                                           \\
HL(16-21)             & -                           & \multicolumn{1}{c|}{-}         & 73.98±0.43                                                                        & 84.79                                                                           \\
HLFT(18)              & -                           & \multicolumn{1}{c|}{-}         & 76.30±0.33                                                                        & 84.69                                                                           \\
HLFT(16-21)           & -                           & \multicolumn{1}{c|}{-}         & 80.24±0.21                                                                        & 84.88                                                                           \\
\multicolumn{1}{c}{-} & \multicolumn{1}{l}{CLIPL}   & \multicolumn{1}{c|}{-}         & 66.22±0.43                                                                        & 60.95                                                                           \\
\multicolumn{1}{c}{-} & \multicolumn{1}{l}{CLIPL-A} & \multicolumn{1}{c|}{-}         & 66.05±0.37                                                                        & 64.59                                                                           \\ \hline
\multicolumn{5}{c}{Multimodal Results}                                                                                                                                                                                                                     \\ \hline
HLFT(16-21)           & \multicolumn{1}{l}{CLIPL}   & \multicolumn{1}{c|}{-}         & 84.34±0.25                                                                        & 86.78                                                                           \\
HLFT(16-21)           & \multicolumn{1}{l}{CLIPL-A} & \multicolumn{1}{c|}{-}         & 84.70±0.19                                                                        & 87.01                                                                           \\
HLFT(16-21)           & -                           & \multicolumn{1}{l|}{Baichuan2} & 79.66±0.13                                                                        & 85.78                                                                           \\
HLFT(16-21)           & \multicolumn{1}{l}{CLIPL-A} & \multicolumn{1}{l|}{Baichuan2} & 83.85±0.35                                                                        & 88.90                                                                           \\ \hline
\end{tabular}}
\vspace{-10pt} 
\end{table}

\vspace{-3pt}
\subsubsection{Unimodal Recognition Results}

We present the experimental results for both unimodal and multimodal approaches in Table \ref{table:2}. For the acoustic modality, HL represents features extracted using the baseline method \cite{mer2024} from the HuBERT-large model, with HL($i$) indicating the feature output from the $i$-th layer. HLFT($i$) indicates the output from the $i$-th layer after fine-tuning the HuBERT-large model with adapters. As shown in the table, the performance of the fine-tuned features demonstrates a notable improvement over the baseline model. The proposed parameter-efficient fine-tuning method achieves superior performance with multi-layer fused features compared to single-layer features. Specifically, HL(16-21) surpasses HL(18), and HLFT(16-21) outperforms HLFT(18), suggesting that complementary information exists among features from different layers, resulting in more robust results. Moreover, the multi-layer fused features obtained through the proposed parameter-efficient fine-tuning method achieve the highest F1 score of 84.88\% on the test set. This method also demonstrates performance improvements of 7.42\% on the Train\&Val set and 1.39\% on the test set compared to the baseline model, validating its effectiveness.

For the visual modality, CLIPL represents the features extracted using the CLIP-Large model, whereas CLIPL-A denotes the features aligned through the proposed visual feature alignment strategy. As shown in Table \ref{table:2}, in comparison to the CLIPL features, the CLIPL-A features exhibit comparable performance on the Train\&Val set and show a 3.64\% improvement on the test set. These results affirm the efficacy of the visual feature alignment strategy in enhancing performance within multimodal emotion recognition tasks.
\vspace{-3pt}
\subsubsection{Multimodal Recognition Results}
We conduct a comprehensive comparison of the multimodal fusion effects, with specific results presented in Table \ref{table:2}. Initially, we fuse the best-performing acoustic modality features, HLFT (16-21), with the visual modality features extracted by CLIP-large. This fusion improves the recognition accuracy from 84.88\% to 86.78\%. Furthermore, using the aligned features extracted by the pre-trained vision MLP, the recognition accuracy further increases to 87.01\%, which provides additional validation for the effectiveness of the feature alignment pre-training method. Similarly, we evaluate the fusion of lexical modality and acoustic modality. When fusing Baichuan2 features with HLFT (16-21) features, the performance on the test set reaches 85.78\%. Ultimately, the fusion of all three modality features results in the highest performance of 88.90\% on the test set. Additionally, the table illustrates that the effect of multimodal fusion on the test set surpasses that of any single modality.

\vspace{-5pt}
\section{Conclusion}
In this study, we propose a multimodal emotion recognition framework for the MER2024-SEMI challenge. Initially, our focus is on fully leveraging acoustic modality features to enhance emotion recognition tasks. We evaluate the performance of different transformer layers of the HuBERT-large model in speech emotion recognition and employ an PEFT method to fine-tune the HuBERT-large model. Subsequently, to enhance the emotional representation of the visual modality, we introduce an unsupervised feature alignment scheme that employs contrastive learning to align visual embeddings with acoustic embeddings. Experimental results validate the effectiveness of the proposed methods, with our approach securing fourth place in the MER2024-SEMI sub-challenge.

\vspace{-5pt}
\section{Acknowledgments}

This work is supported by the National Natural Science Foundation of China (grant 62236006, grant 62306172), the Key Research and Development Program of Shaanxi (No. 2022ZDLGY06-03).

%
\bibliographystyle{ACM-Reference-Format}
\balance
\bibliography{mybib}










\end{document}